\documentclass[twocolumn,showpacs,preprintnumbers,amsmath,amssymb,prl]{revtex4-1}

\usepackage{graphicx}
\usepackage{dcolumn}
\usepackage{bm}
\usepackage{amsmath}
\usepackage{amssymb}

\usepackage{color}
\usepackage{braket}
\usepackage{lipsum}

\begin{document}

\newcommand{\note}[1]{\textbf{#1}}

\title{Displacement of propagating squeezed microwave states}
\author{\firstname{Kirill G.} \surname{Fedorov$^{1}$}}
\email{kirill.fedorov@wmi.badw.de}
\author{\firstname{L.}~\surname{Zhong$^{1,2,3}$}}
\author{\firstname{S.}~\surname{Pogorzalek$^{1,2}$}}
\author{\firstname{P.}~\surname{Eder$^{1,2,3}$}}
\author{\firstname{M.}~\surname{Fischer$^{1,2}$}}
\author{\firstname{J.}~\surname{Goetz$^{1,2}$}}
\author{\firstname{E.}~\surname{Xie$^{1,2,3}$}}
\author{\firstname{F.}~\surname{Wulschner$^{1,2}$}}
\author{\firstname{K.}~\surname{Inomata$^{6}$}}
\author{\firstname{T.}~\surname{Yamamoto$^{7}$}}
\author{\firstname{Y.}~\surname{Nakamura$^{6,8}$}}
\author{\firstname{R.}~\surname{Di Candia$^{4}$}}
\author{\firstname{U.}~\surname{Las Heras$^{4}$}}
\author{\firstname{M.}~\surname{Sanz$^{4}$}}
\author{\firstname{E.}~\surname{Solano$^{2,4,5}$}}
\author{\firstname{E. P.}~\surname{Menzel$^{1}$}}
\author{\firstname{F.}~\surname{Deppe$^{1,2,3}$}}
\author{\firstname{A.}~\surname{Marx$^{1}$}}
\author{\firstname{R.}~\surname{Gross$^{1,2,3}$}}

\affiliation
{
$^{1}$ Walther-Mei{\ss}ner-Institut, Bayerische Akademie der Wissenschaften, D-85748 Garching, Germany \\
$^{2}$ Physik-Department, Technische Universit\"{a}t M\"{u}nchen, D-85748 Garching, Germany \\
$^{3}$ Nanosystems Initiative Munich (NIM), Schellingstra{\ss}e 4, 80799 M\"{u}nchen, Germany \\
$^{4}$ Department of Physical Chemistry, University of the Basque Country UPV/EHU, Apartado 644, E-48080 Bilbao, Spain \\
$^{5}$ IKERBASQUE, Basque Foundation for Science, Maria Diaz de Haro 3, 48013 Bilbao, Spain \\
$^{6}$ RIKEN Center for Emergent Matter Science (CEMS), Wako, Saitama 351-0198, Japan \\
$^{7}$ NEC Smart Energy Research Laboratories, Tsukuba, Ibaraki 305-8501, Japan \\
$^{8}$ Research Center for Advanced Science and Technology (RCAST), The University of Tokyo, Meguro-ku, Tokyo 153-8904, Japan
}

\date{\today}

\begin{abstract}
Displacement of propagating quantum states of light is a fundamental operation for quantum communication. It enables fundamental studies on macroscopic quantum coherence and plays an important role in quantum teleportation protocols with continuous variables. In our experiments we have successfully implemented this operation for propagating squeezed microwave states. We demonstrate that, even for strong displacement amplitudes, there is no degradation of the squeezing level in the reconstructed quantum states. Furthermore, we confirm that path entanglement generated by using displaced squeezed states stays constant over a wide range of the displacement power.
\end{abstract}



\maketitle

Propagating quantum microwave signals in the form of squeezed states are promising candidates for information processing and communication tasks in superconducting quantum networks. With respect to frequency range and material technology for transmission lines, quantum microwaves can directly interact with information processing platforms based on superconducting quantum circuits (SQC). Thus, additional circuitry for frequency interconversion and a possible technology mismatch are avoided \cite{QStateTomog2011, DualPath2010, PathEnt2012, DPMethods2014, TwoModeSq2011, TomoItinMw2011}. Long coherence times demonstrated for microwave photons in superconducting resonators \cite{SCRes1,SCRes2} translate into distances in the range of kilometers for propagating waves in superconducting environments. Therefore, SQC in combination with propagating microwaves could provide the basis for the implementation of a short- to medium-range quantum communication and information processing. One of the cornerstones of quantum communication is the paradigm of quantum teleportation which allows one to faithfully transmit an unknown quantum state between two spatially-separated parties using a quantum entangled pair and a classical communication channel. Various experiments in the past have demonstrated the feasibility of this idea in different quantum technologies such as quantum optics \cite{CVQuanTele1998,DVQuanTele1997}, superconducting quantum circuits \cite{DisQuanTele2013}, and atomic systems \cite{Riebe1}, among others.

Fundamental operations needed to implement the quantum teleportation protocol with continuous variables \cite{DiCandiatel} include the generation of two-mode squeezing, quadrature measurements, and a conditional displacement. While there are experimental advances for the first two operations \cite{PathEnt2012,TwoModeSq2011,EntRad2012}, a controllable displacement has not yet been demonstrated in the microwave regime. It is important to note that displacement belongs to the universal set of quantum gates required for quantum information processing with continuous variables \cite{BraunRev2005,DiscContQInf2015}. Moreover, from a more fundamental point of view, displacement can allow one to study very general limits of quantum entanglement and coherence \cite{DispEntang2013}. Therefore, it is important to investigate capabilities and possible limitations of the displacement operation applied to non-classical propagating microwave states.
\begin{figure}
        \begin{center}
        \includegraphics[width=\linewidth,angle=0,clip]{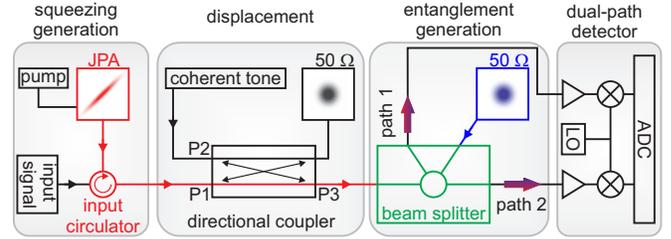}
        \end{center}
    \caption{Circuit schematic for the displacement of propagating squeezed microwave states. The displacement amplitude and phase are controlled by a coherent microwave tone. The 50:50 microwave beam splitter is realized as a 180\,$^\circ$ cryogenic hybrid ring. It superimposes the squeezed state and a vacuum state from a cold $50$\,$\Omega$ load, thus, generating path-entangled output states (blue-red arrows). State detection is realized with the dual-path detector \cite{DualPath2010,PathEnt2012,DPMethods2014} based on the cross-correlation heterodyne scheme, where LO is a local oscillator and ADC denotes an analog-to-digital converter.}
   \label{fig1}
\end{figure}

In this Letter, we experimentally confirm the feasibility of the displacement operation on propagating squeezed microwaves. Specifically, we demonstrate that both a single mode squeezing and a frequency-degenerate continuous variable path entanglement remain unchanged over the range of $30$ decibel in the displacement power. These results are a very encouraging and a crucial step on the way towards quantum microwave teleportation with continuous variables \cite{DiCandiatel}.

\begin{figure}
        \begin{center}
        \includegraphics[width=\linewidth,angle=0,clip]{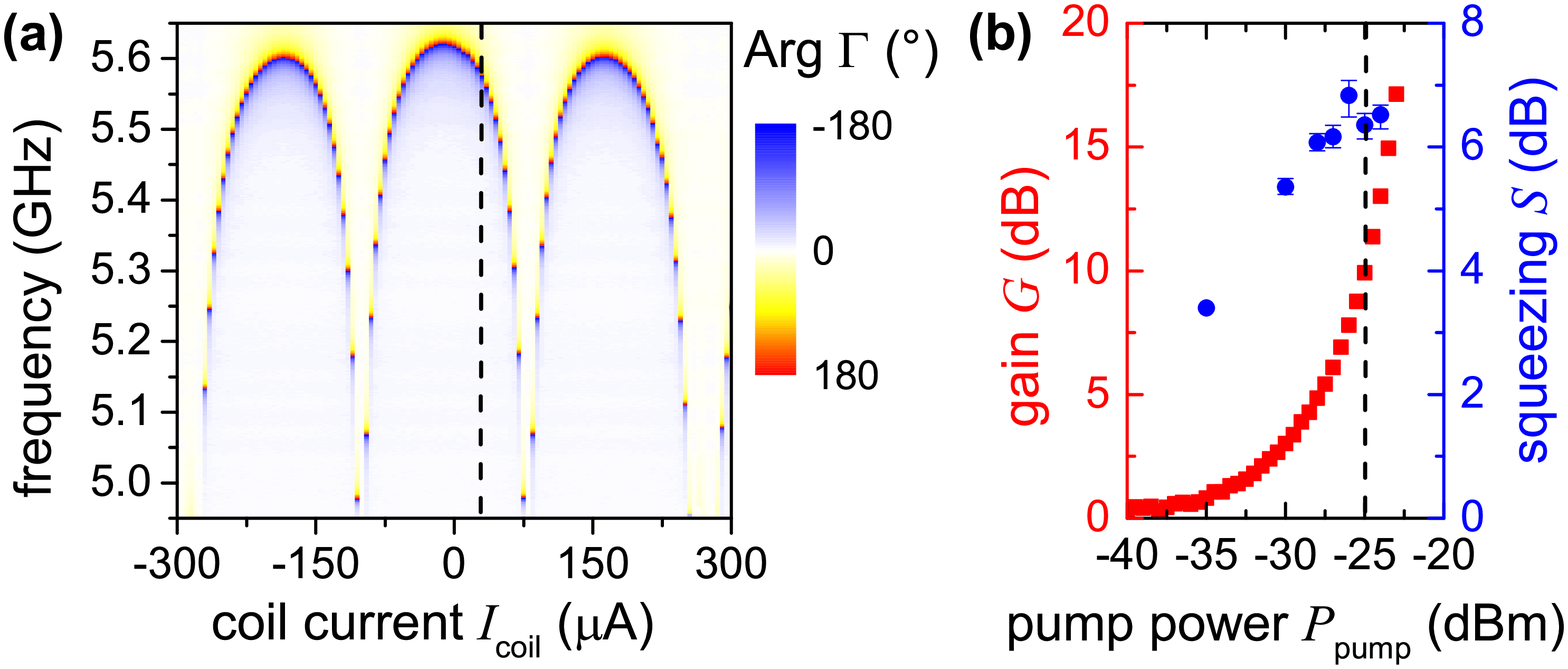}
        \end{center}
    \caption{\textbf{(a)} Spectroscopy of the JPA measured using a vector network analyzer. Phase of the reflected signal $\Gamma$ is plotted as a function of the
    coil current $I_{\rm{coil}}$ generating a magnetic flux in the dc-SQUID loop. The dashed black line marks the working point $f_{\rm{JPA}}\,=\,5.573\,$GHz
    corresponding to $I_{\rm{coil}}\,=\,+30\,\mu$A. \textbf{(b)} Measurements of the non-degenerate gain $G$ (squares) and the squeezing level $S$ (circles) versus the pump power $P_{\rm{pump}}$ incident at the input of the sample holder. The pump frequency is $f_{\rm{pump}} \,=\, 2 f_{\rm{JPA}}$. The dashed black line marks the working point for the pump ($P_{\rm{pump}} \,{=}\, - 25\,$dBm) used for further displacement measurements.}
    \label{fig2}
\end{figure}
Figure\,\ref{fig1} shows a circuit schematic of our experimental setup. We use a flux-driven Josephson parametric amplifier (JPA) for the generation of squeezed microwave states \cite{FD-JPA2008,JPA-Ch2013}. The JPA consists of a quarter-wavelength coplanar waveguide resonator shunted to ground with a dc-SQUID. The dc-SQUID is inductively coupled to an on-chip antenna used for the application of a pump signal. The strong pump tone allows one to modulate the Josephson inductance of the dc-SQUID at twice the JPA frequency $f_{\rm{pump}} = 2 f_{\rm{JPA}}$, thus, fulfilling the condition for parametric amplification \cite{JPA-Ch2013}. Additional details, including  the sample layout, are provided in Ref.\,\onlinecite{SuppM}. The JPA is placed in a magnetically shielded sample-holder (we use a combination of cryoperm and superconducting shields) inside a custom-made dry dilution refrigerator. During all experiments the JPA temperature is stabilized at $50\,$mK. An input circulator allows
us to separate incoming and outgoing signals, as required for the JPA operation. We calibrate our setup via a temperature sweep of a cold $30\,$dB microwave attenuator, which produces a thermal state with a well-known photon distribution and, therefore, allows us to determine a total gain of the cross-correlation detector \emph{in situ} \cite{PlanckSpec2010,PathEnt2012,JPA-Ch2013}.

The task of the JPA is to perform squeezing of the incident vacuum state, i.e., $\hat{S}(\xi) |0\rangle$, where $\hat{S}(\xi)\,{=}\, \exp(\frac{1}{2}\xi^* \hat{a}^2\,{-}\,\frac{1}{2}\xi (\hat{a}^\dagger)^2)$, and $\xi = r e^{i\phi}$ is a complex squeezing amplitude.  Here, the phase $\phi$ determines the squeezed quadrature, while the squeezing factor $r$ parameterises the amount of squeezing. An experimental realization of the squeezing operator requires a large nonlinearity. Figure\,\ref{fig2} illustrates such a nonlinear flux dependence (measured as the phase response of a reflected input signal from the sample) of the JPA resonance frequency. It also provides information about the nondegenerate gain as a function of the pump power $P_{\rm{pump}}$ at a chosen working point $I_{\rm{coil}} = +30\,\mu$A. For all subsequent measurements,
 we use a fixed pump power of $P_{\rm{pump}} = -25\,$dBm, which corresponds to a nondegenerate gain of $G \simeq 9.9\,$dB at the JPA operation frequency $f_{\rm{JPA}} = 5.573\,$GHz. Higher pump powers allow us to reach even higher gains and, accordingly, higher squeezing levels, but lead to unsustainable heating of the dilution refrigerator.

\begin{figure*}
    \centering
    \includegraphics[width=\linewidth,angle=0,clip]{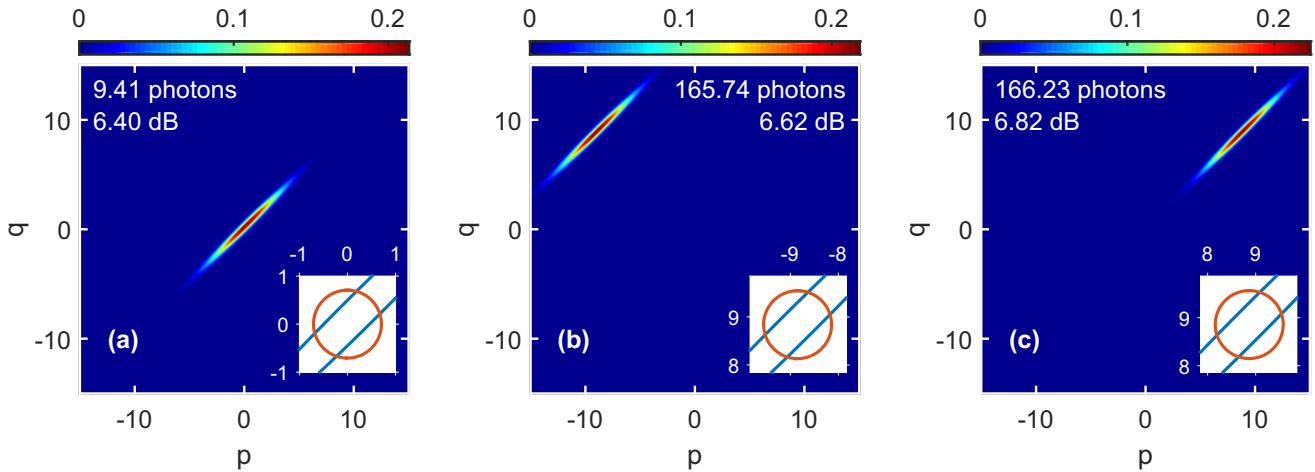}
    \caption{Reconstructed squeezed states at the input of the hybrid ring. The quantities $p$ and $q$ are dimensionless variables spanning the phase space.
    The squeezing angle is stabilized at $\gamma = 45$ $^{\circ}$. The color code represents the Wigner function of \textbf{(a)} a squeezed vacuum state, \textbf{(b)} a displaced squeezed vacuum state with a displacement angle of $\theta = 135$ $^{\circ}$, \textbf{(c)} a displaced squeezed vacuum state with a displacement angle of $\theta = 45$ $^{\circ}$. Each measurement is averaged over $1.5 \times 10^9$ samples. The total photon numbers and squeezing levels are indicated in each panel. The insets show $1/e$ contours for the ideal vacuum (red), and experimental squeezed states (blue).}
    \label{fig3}
\end{figure*}
In order to implement the displacement operation $\hat{D}(\alpha)=\exp(\alpha\hat a^\dag-\alpha^* \hat a)$ on propagating squeezed states, we use a cryogenic directional coupler designed according to Ref.\,\onlinecite{DispOper1996}. The complex displacement parameter $\alpha$ is controlled via the amplitude and the phase of a strong coherent signal incident at the coupling port of the directional coupler (see Fig.\,\ref{fig1}). The propagating squeezed state from the JPA is applied to the input port P1 of the coupler. The coupling between the transmitted port P3 and the displacement port P2 is $-19.5\,$dB in the vicinity of $f_{\rm{JPA}}$. The insertion loss of the coupler is $\kappa \simeq -0.18\,$dB (between ports P1 and P3). Here, the directional coupler is analogous to a $99 \%$ reflective beam splitter, in which the displacement signal is weakly coupled to the incident light.

In order to reconstruct propagating displaced squeezed states, we apply the dual-path reconstruction scheme \cite{DualPath2010,PathEnt2012}, where we first equally
distribute the incoming signal along two paths with a hybrid ring. Both outputs are independently amplified with a chain of cryogenic and room temperature rf-amplifiers, downconverted in a two-stage process, and finally used for cross-correlation measurements. In the analog stage, a local oscillator (LO, see Fig.\,\ref{fig1}) defines a phase reference for detected signals. An important modification (in comparison with Refs.\,\onlinecite{DualPath2010, PathEnt2012}) of our setup consists in using image reject mixers, which allow us to filter one of the signal sidebands already in the analog part, reducing the number of digitizer channels from 4 to 2. The path-entangled signals are mixed with the LO and produce intermediate frequency (IF) signals at $f_{\rm{IF}} = f_{\rm{LO}} - f_{\rm{JPA}} = 11$ MHz. Then, the IF signals are digitized by the ADC card and processed by a measurement program which extracts orthogonal quadratures $I_{1,2}$ and $Q_{1,2}$ by performing down-conversion and digital filtering. Extended experimental schematics may be found in Ref.\,\onlinecite{SuppM}. The quadrature components are used to calculate correlation moments $\langle I_1^n I_2^m Q_1^k Q_2^l \rangle$ up to the fourth order, i.e., $n + m + k + l \le 4$ for $n,m,k,l\,{\in}\,\mathbb{N}$. In this way, we can retrieve all the moments of the annihilation and creation operators, $\hat{a}$ and $\hat{a}^{\dagger}$, of the signal and the noise modes by using the beam splitter relations and the independence of the noise contributions from the two detection paths. Finally, based on the operator moments
$\braket{(\hat{a}^{\dagger})^n \hat{a}^m}$, we calculate the covariance matrix $\vec{\sigma}$ of the signal mode, and reconstruct the Wigner function of the
propagating signal incident at the input of the hybrid ring. We verify that the reconstructed states comply with the Heisenberg principle for the $2$-nd and the $4$-th order moments $\braket{(\hat{a}^{\dagger})^n \hat{a}^m}$ \cite{QMcrMoments2011}, and with a Gaussianity criterion based on cumulants \cite{PathEnt2012,Cumul1,QLAmpEnt2014}.

\begin{figure}
        \begin{center}
        \includegraphics[width=\linewidth,angle=0,clip]{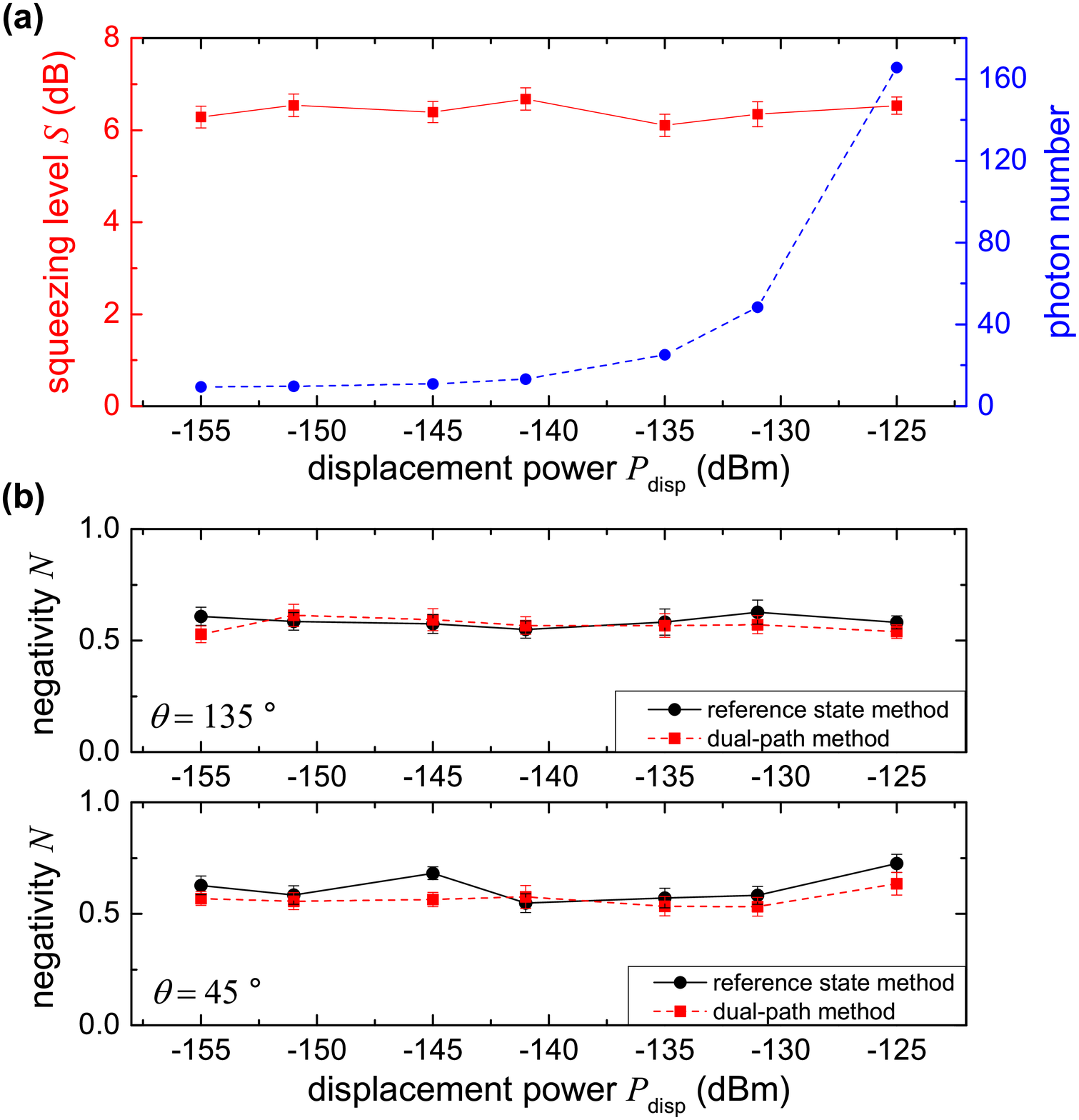}
        \end{center}
    \caption{\textbf{(a)} Squeezing level $S$ (squares) and photon number (circles) of displaced squeezed state as a function of displacement power incident at the coupled port of the directional coupler. \textbf{(b)} Measured negativity as a function of displacement power for two orthogonal displacement angles (also used in Fig.\,\ref{fig3}) $\theta = 135\,^{\circ}$, and $\theta = 45\,^{\circ}$. One data point (symbols) is averaged over $7.5 \times 10^{10}$ samples.
    Lines are guides to the eye. The error bars are of statistical nature.}
    \label{fig4}
\end{figure}

We characterize the squeezing level of the reconstructed quantum state in decibels as $S\,{=}\,{-}10\,\log_{10} [(\Delta X_{\rm{sq}})^2 / 0.25]$, where $(\Delta X_{\rm{sq}})^2$ is the variance of the squeezed quadrature and the chosen vacuum reference is $(\Delta X_{\rm{vac}})^2\equiv0.25$. A state is squeezed below the
vacuum level when $(\Delta X_{\rm{sq}})^2 < 0.25$. We define the displacement angle $\theta$ as the angle between the displacement direction and the $p$-axis, and the squeezing angle $\gamma = − \phi/2$ as the angle between the antisqueezed quadrature and the $p$-axis. The angles are controlled experimentally by the phases of the coherent signals of the pump and the displacement tones with respect to the LO (cf. Fig.\,\ref{fig1}). Figure\,\ref{fig3} illustrates experimentally reconstructed Wigner functions of a squeezed vacuum and displaced squeezed vacua. The results show that the directional coupler allows us to displace the propagating squeezed state with a high degree of control over the magnitude and the phase of the displacement parameter $\alpha$. Even for a large displacement powers up to hundreds of photons, the resulting squeezing level remains approximately the same as for the undisplaced state. Notice that the photon number is related to a measurement bandwidth of $400$\,kHz, defined by a digital low-pass filter in the data processing. Systematic measurements of both the squeezing level and the total
photon number versus the displacement power $P_{\rm{disp}}$ are shown in Fig.\,\ref{fig4}\,(a). There, we show that the resulting squeezing level for $P_{\rm{disp}}\,\le\,-125\,$dBm is the same as for the undisplaced state, $S\,\simeq\,6.4\,$dB, within error bars. For larger displacement powers, $P_{\rm{disp}}\,{>}\,-125\,$dBm (data not shown), we observe an increasing degradation of squeezing. Based on numerical simulations, we attribute this observation to a limited calibration precision of the cross-correlation gain which distorts the reconstruction of states with a very high photon number. It is important to note here, that the demonstrated range of the controlled displacement of propagating squeezed states is already sufficient for many quantum communication protocols with
continuous-variable microwaves including quantum teleportation \cite{BraunRev2005,DiCandiatel}.

As it was confirmed in previous experiments \cite{PathEnt2012}, the squeezing at the hybrid ring input is equivalent to a path entanglement of the output beams. Therefore, from a fundamental point of view, one must check whether the amount of bipartite path entanglement between the outputs of the entangling hybrid ring depends on the displacement power. To this end, we measure the negativity $N$.
For Gaussian states \cite{Negat2005, PathEnt2012}, it is defined as $N \,{\equiv}\, \max \left\{ 0, (1-\nu)/(2\nu) \right\}$, where
$\nu \,{\equiv}\, ((\Delta(\vec{\sigma})-(\Delta^2(\vec{\sigma}) - 4 \det \vec{\sigma})^{0.5})/2)^{0.5}$ and $\Delta(\vec{\sigma}) = \det \vec{\alpha} + \det \vec{\beta} - 2 \det \vec{\gamma}$. The condition $N > 0$ witnesses the presence of entanglement for a general quantum state. The covariance matrix $\vec{\sigma}$ of the displaced squeezed state $\hat{D}(\alpha) \hat{S}(\xi) |0\rangle$ incident to the hybrid ring is given by
\begin{equation}
\label{covmat1}
\vec{\sigma}=
  \begin{pmatrix}
    \vec{\alpha} & \vec{\gamma}  \\
    \vec{\gamma}^T & \vec{\beta}
  \end{pmatrix}, \,\,\,
\vec{\alpha}= \frac{1}{2}
  \begin{pmatrix}
    A_{-}+1 & B  \\
    B & A_{+}+1
  \end{pmatrix},
\end{equation}
\begin{equation}
\label{covmat2}
\vec{\beta}= \vec{\alpha}, \,\,\,
\vec{\gamma}=
\frac{1}{2}
  \begin{pmatrix}
    A_{-} - 1 & B \\
    B & A_{+} - 1
  \end{pmatrix},
\end{equation}
where $A_{\pm}\,\equiv\, e^{\pm 2r} \cos^2(\phi/2) + e^{\pm 2r} \sin^2(\phi/2)$, and $B \,\equiv\, -\sinh(2r) \sin(\phi)$.
We immediately see that the covariance matrix does not explicitly depend on the displacement parameter $\alpha$. Hence, the amount of path entanglement should not change regardless of the magnitude or the phase of the displacement. Nevertheless, in usual decoherence environments, one would expect a transition between an entangled quantum behaviour and a classical separable description for large displacements.

Figures\,\ref{fig4}(b) depicts the negativity $N$ versus the displacement power $P_{\rm{disp}}$ for two different displacement angles $\theta$. Here, we determine the negativity with two methods: the dual-path method (DPM) and the reference state method (RSM) \cite{DualPath2010, PathEnt2012, DPMethods2014}. The DPM estimates the negativity based on the beam splitter model and is mathematically equivalent to the Wigner function reconstruction of the state incident at
the \emph{input} of the hybrid ring. In contrast, the RSM is an independent experiment and reconstructs the moments of the \emph{output} state using a calibration
against a known reference signal which, in our case, is the two-mode vacuum. Thus, the RSM provides a direct evidence of the entanglement between two spatially
separated modes. In Fig.\,\ref{fig4}(b), we show the results for both methods, proving that they coincide in our experiments. The good agreement between both methods confirms the high degree of control and understanding we have on our experiment. Both methods confirm the independence of the path entanglement from the displacement amplitude.

In conclusion, we have experimentally studied the displacement of propagating squeezed microwave states. We have shown that even states displaced by hundreds of photons do not lose their original squeezing level. The implemented displacement mechanism is an important step towards quantum communication and
information processing experiments with continuous-variable quantum microwaves \cite{BraunRev2005,DiCandiatel}. Furthermore, we have applied this operation to general studies of the path entanglement generated from displaced squeezed microwaves via a beam splitter. We have demonstrated that the path entanglement is
preserved over a wide range of displacement power. Finally, we interpret our results as an experimental spotlight on a more general issue in quantum mechanics:
the quantum properties of large photonic states with hundreds of photons can be associated with a relatively small subset of these photons.  In our specific case, the squeezed photons translate into entanglement after the beam splitter while the displacement photons do not.

We acknowledge support by the German Research Foundation through SFB 631 and FE 1564/1-1, the EU projects PROMISCE and SCALEQIT, Spanish MINECO FIS2012-36673-C03-02, UPV/EHU Project EHUA14/04, UPV/EHU PhD grant, Basque Government IT472-10, Elite Network of Bavaria through the program ExQM, the Project for Developing Innovation Systems of MEXT, and the NICT Commissioned Research, and ImPACT Program of Council for Science, Technology and Innovation. E.S. acknowledges support from
a TUM August-Wilhelm Scheer Visiting Professorship and hospitality of Walther-Mei{\ss}ner-Institut and TUM Institute for Advanced Study. We would like to thank K. Kusuyama for assistance with part of the JPA fabrication. K.G.F. would like to thank O. Buisson for useful discussions.


\begin{thebibliography}{99}

\bibitem{QStateTomog2011} F. Mallet, M. A. Castellanos-Beltran, H. S. Ku, S. Glancy, E. Knill, K. D. Irwin, G. C. Hilton, L. R. Vale, and K. W. Lehnert, Phys. Rev. Lett. \textbf{106}, 220502 (2011).
\bibitem{DualPath2010} E. P. Menzel, M. Mariantoni, F. Deppe, M. A. Araque Caballero, A. Baust, T. Niemczyk, E. Hoffmann, A.Marx, E. Solano, and R. Gross,  Phys. Rev. Lett. \textbf{105}, 100401 (2010).
\bibitem{PathEnt2012} E. P. Menzel, R. Di Candia, F. Deppe, P. Eder, L. Zhong, M. Ihmig, M. Haeberlein, A. Baust, E. Hoffmann, D. Ballester, K. Inomata, T. Yamamoto, Y. Nakamura, E. Solano, A. Marx, and R. Gross, Phys. Rev. Lett. \textbf{109}, 250502 (2012).
\bibitem{DPMethods2014} R. Di Candia, E. P. Menzel, L. Zhong, F. Deppe, A. Marx, R. Gross, and E. Solano, New J. Phys. 16, 015001 (2014).
\bibitem{TwoModeSq2011} C. Eichler, D. Bozyigit, C. Lang, M. Baur, L. Steffen, J. M. Fink, S. Filipp, and A. Wallraff,  Phys. Rev. Lett. \textbf{107}, 113601 (2011).
\bibitem{TomoItinMw2011} C. Eichler, D. Bozyigit, C. Lang, L. Steffen, J. Fink, and A. Wallraff, Phys. Rev. Lett. \textbf{106}, 220503 (2011).
\bibitem{SCRes1} A. Megrant, C. Neill, R. Barends, B. Chiaro, Yu Chen, L. Feigl, J. Kelly, E. Lucero, M. Mariantoni, P. J. J. O’Malley, D. Sank, A. Vainsencher, J. Wenner, T. C. White, Y. Yin, J. Zhao, C. J. Palmstrøm, J. M. Martinis, and A. N. Cleland, Appl. Phys. Lett. \textbf{100}, 113510 (2012).
\bibitem{SCRes2} M. Reagor, H. Paik, G. Catelani, L. Sun, C. Axline, E. Holland, I. M. Pop, N. A. Masluk, T. Brecht, L. Frunzio, M. H. Devoret, L. Glazman, and R. J. Schoelkopf, Appl. Phys. Lett. \textbf{102}, 192604 (2013).
\bibitem{CVQuanTele1998} A. Furusawa, J. L. Sorensen, S. L. Braunstein, C. A. Fuchs, H. J. Kimble, and E. S. Polzik, Science \textbf{282}, 706 (1998).
\bibitem{DVQuanTele1997} D. Bouwmeester, J. Pan, K. Mattle, M. Eibl, H. Weinfurter, and A. Zeilinger, Nature \textbf{390}, 575 (1997).
\bibitem{DisQuanTele2013} L. Steffen, Y. Salathe, M. Oppliger, P. Kurpiers, M. Baur, C. Lang, C. Eichler, G. Puebla-Hellmann, A. Fedorov, and A. Wallraff, Nature \textbf{500}, 319 (2013).
\bibitem{Riebe1} M. Riebe, H. H\"affner, C. F. Roos, W. H\"ansel, M. Ruth, J. Benhelm, G. P. T. Lancaster, T. W. K\"orber, C. Becher, F. Schimdt-Kaler, D. F. V. James, and R. Blatt, Nature \textbf{429}, 734 (2004).
\bibitem{DiCandiatel} R. Di Candia, K. G. Fedorov, L. Zhong, S. Felicetti, E. P. Menzel, M. Sanz, F. Deppe, A. Marx, R. Gross, and E. Solano, EPJ Quantum Technology \textbf{2}, 25 (2015).
\bibitem{EntRad2012} E. Flurin, N. Roch, F. Mallet, M. H. Devoret, and B. Huard, Phys. Rev. Lett. \textbf{109}, 183901 (2012).
\bibitem{BraunRev2005} S. L. Braunstein, and P. van Loock, Rev. Mod. Phys. \textbf{77}, 513 (2005).
\bibitem{DiscContQInf2015} U. L. Andersen, J. S. Neergaard-Nielsen, P. van Loock, and A. Furusawa, Nat. Phys. \textbf{11}, 713 (2015).
\bibitem{DispEntang2013} N. Bruno, A. Martin, P. Sekatski, N. Sangouard, R. T. Thew, and N. Gisin, Nat. Phys. \textbf{9}, 545 (2013).
\bibitem{FD-JPA2008} T. Yamamoto, K. Inomata, M. Watanabe, K. Matsuba, T. Miyazaki, W. D. Oliver, Y. Nakamura, and J. S. Tsai, Appl. Phys. Lett. \textbf{93}, 042510 (2008).
\bibitem{JPA-Ch2013} L. Zhong, E. P. Menzel, R. Di Candia, P. Eder, M. Ihmig, A. Baust, M. Haeberlein, E. Hoffmann, K. Inomata, T. Yamamoto, Y. Nakamura, E. Solano, F. Deppe, A. Marx and R. Gross, New J. Phys. \textbf{15}, 125013 (2013).
\bibitem{SuppM} See Supplemental Material at [url].
\bibitem{PlanckSpec2010} M. Mariantoni, F. Deppe, M.A. Araque, A. Baust, T. Niemczyk, E. Hoffmann, E. Solano, A. Marx, and R. Gross, Phys. Rev. Lett. \textbf{105}, 133601 (2010).
\bibitem{DispOper1996} M.G.A. Paris, Phys. Lett. A \textbf{217}, 78 (1996).
\bibitem{QMcrMoments2011} S. N. Filippov, and V. I. Man'ko, Phys. Rev \textbf{84}, 033827 (2011).
\bibitem{Cumul1} R. Schack and A. Schenzle, Phys. Rev. A \textbf{41}, 3847 (1990).
\bibitem{QLAmpEnt2014} C. Eichler, Y. Salathe, J. Mlynek, S. Schmidt, and A. Wallraff \textbf{113}, Phys. Rev. Lett. 110502 (2014).
\bibitem{Negat2005} G. Adesso and F. Illuminati, Phys. Rev. A \textbf{72}, 032334 (2005).

\end{thebibliography}
\end{document}